\documentclass[twocolumn,pre,amsmath,amssymb,floatfix,preprintnumbers]{revtex4-1}

\usepackage{graphicx}% Include figure files
\usepackage{dcolumn}% Align table columns on decimal point
\usepackage{bm}% bold math
\usepackage[utf8]{inputenc}
\newcommand{\hh}{\{h\}}

\newcommand{\pp}{\{p\}}
\newcommand{\ppp}{\{p^*\}}
\newcommand{\nn}{\{n\}}

\begin{document}

\preprint{Thermodynamics 2.0 Meeting, June 22-24, 2020}

\title{Thermodynamics and the Evolution of Stochastic Populations}

\author{Themis Matsoukas}
\email{txm11@psu.edu}
\affiliation{Department of Chemical Engineering, The Pennsylvania State University, University Park, PA 16802.}

\date{June 20, 2020}
%--------------------------------------------------------------------------*
\begin{abstract}
The appeal of thermodynamics to problems outside physics is undeniable, as is the growing recognition of its apparent universality, yet in the absence of a rigorous formalism divorced from the peculiarities of molecular systems all attempts to generalize thermodynamics remain qualitative and heuristic at best. In this paper we formulate a probabilistic theory of thermodynamics and and set the basis for its application to generic stochastic populations. \end{abstract}
%--------------------------------------------------------------------------*
%%\pacs{PACKS}
%%\keywords{keywords}
\maketitle
%--------------------------------------------------------------------------*
\section{WHAT IS THERMODYNAMICS?}

There is growing recognition that thermodynamics is a universal science but what \textit{is}  thermodynamics? The typical answer is some variation of ``thermodynamics is the study of the equilibrium properties of matter.'' Undoubtedly the study of the equilibrium state of matter falls within the scope of thermodynamics, clearly though, such statement is far too narrow to serve as a definition of thermodynamics, far too narrow to justify its appeal in areas \textit{outside} physics. The answer describes one particular problem (albeit a very important one) that thermodynamics can tackle, but tells little about what thermodynamics is capable of. \textit{What is thermodynamics}?

Until Boltzmann and Gibbs thermodynamics was about heat and work and its central aim was to determine the maximum amount of work that can be extracted from heat in a cycle. This is a variational problem and its solution was found via a variational principle,  the maximization of entropy. With Boltzmann and Gibbs \cite{Gibbs:reprint}, thermodynamics entered the realm of the probabilistic world  --while still footed in the physical world of mechanical particles, atoms and molecules. Entropy now has a probabilistic interpretation, it is a functional of the probability distribution $p_i$ of microstates,
\begin{equation}
\label{S_functional}
   S[\pp]= - \sum_i p_i \log p_i .
\end{equation}
The equilibrium distribution $p^*_i$ is identified among all feasible distributions as the one that maximizes this functional:
\begin{equation}
\label{S_variational}
   S[\ppp] \geq S[p] . 
\end{equation}
A feasible distribution is one that satisfies the macroscopic constraints that define the system. For a system with fixed mean energy $\bar E$, volume $V$ and number of particles $N$ this means any distribution with mean energy $\bar E$, i.e., any distribution that satisfies 
\begin{equation}
\label{barE}
   \sum_i E_i p_i = \bar E . 
\end{equation}
The solution to this variational problem is the canonical probability of microstate,
\begin{equation}
\label{p_canonical}
   p^*_i = \frac{e^{-\beta E_i}}{Q(\beta,V,N)}, 
\end{equation}
$Q$ is the canonical partition function and $\beta$ is the inverse temperature. With minimal additional calculus $\beta$ and $Q$ are expressed in terms of the microcanonical partition function $\Omega(E,V,N)$, 
\begin{equation}
\label{thermo_1}
   \beta = \frac{\partial\log\Omega}{\partial E};\quad
   \log Q = \log \Omega - \bar E\frac{\partial\log\Omega}{\partial E} ; 
\end{equation}
and similarly $\bar E$ is expressed in terms of the canonical partition function
\begin{equation}
\label{thermo_2}
   \bar E = -\frac{\partial\log Q}{\partial\beta} . 
\end{equation}
Equations (\ref{S_functional}) through (\ref{thermo_2}) cast the variational problem in terms of a \textit{probability space}, the space of all probability distributions that are feasible under the macroscopic specification of state. 
These equations bear only indirect relationship to the physical laws that govern that particles that produce the microstates in question via $E_i$.  It should not be a surprise  that thermodynamics survived the quantum revolution.  Quantum mechanics changes the calculation of the energy $E_i$ of microstate but leaves the entire network of Eqs.\ (\ref{S_variational})--(\ref{thermo_2}) intact. The hint is clear: thermodynamics is \textit{not} a physical theory. 

The door that allowed thermodynamics to escape the confines of physics was unlocked by Shannon, who arrived at the entropy functional, the same functional as that of Boltzmann and Gibbs, by methods that have nothing to do with physical particles or laws of motion \cite{Shannon:BSTJ}. Even so the door remained closed until Jaynes pushed it open. Jaynes conjectured that thermodynamics is about inference, not physics: 
\begin{quote}\em \small
[...] we may have now reached a state where statistical mechanics is no longer dependent on physical hypotheses, but may become simply an example of statistical inference \cite{Jaynes:PR57}.
\end{quote}
With this bold statement the toolbox of thermodynamics was made available to anyone who can  use it. 
During the more than 60 years since Jaynes's pronouncement thermodynamics has been steadily moving closer to achieving the status of universal language with applications in  ecology \cite{Harte:E08}, epidemics \cite{Durrett:SR99}, neuroscience \cite{Timme:E18}, financial markets \cite{Voit:05}, and in the study of complexity in general. But what are the rules of this universal science? We have no comprehensive theory of generalized thermodynamics and no guidance on how to connect a particular system to thermodynamics. One is largely left to construct analogies to molecular systems and hope they hold. 

Here we attempt to decipher the universal grammar of thermodynamics. First, we construct a mathematical formalism based on variational calculus that generates the entire network of mathematical thermodynamics, Eqs.\ (\ref{S_functional}) through (\ref{thermo_2}), via a variational construction that assigns probabilities to a set of probability distributions and seeks the most probable distribution in the set. The most probable distribution is identified by maximizing its probability and this leads to the maximization of a functional analogous to but different from entropy, in fact, a generalized (negative) free energy. We show that any probability distribution can be expressed in the formalism of thermodynamics and illustrate how these ideas can be applied to a generic population that evolves under a stochastic process. The paper builds on work recently published in Refs.\ \cite{Matsoukas:springer_2019} and \cite{Matsoukas:E19}. 

%--------------------------------------------------------------------------* 
\section{THERMODYNAMICS BEYOND MOLECULES}
\subsection{Sampling Statistics}
We start with a discrete random variable $X=i$, $i=1,2\cdots$ with probability distribution $P(X=i) = h_i$. We collect a random sample of $N$ points and count the number $n_i$ of samples with $X=i$. The (intensive) empirical distribution of the sample is $p_i = n_i/N$ and represents a guess of the true distribution $h_i$. The probability to obtain a particular (extensive) distribution $\nn=N \hh $ is
\begin{equation}
   P(\nn) = P(\hh ) 
   = N! \prod_{i=1}^N \frac{h_i^{n_i}}{n_i!} , 
\end{equation}
and its log is
\begin{equation}
   \log P(\hh ) = 
   -N \sum_i p_i \log\frac{p_i}{h_i} .
\end{equation}
We recognize the result as the Kullback-Leibler divergence \cite{Kullback:AMS51} (or relative entropy). This functional, as is well known, is maximized by $p^*_i=h_i$, a result that makes intuitive sense: in a random sample taken from some distribution the most likely distribution to materialize in the sample is the distribution that is being sampled, even though other distributions are possible. When $N$ is large (thermodynamic limit), $p_i$ is overwhelmingly more probable than any other distribution.

We now bias the sampling process so that we may obtain a different distribution  as the most probable distribution from the one that is being sampled. We bias the process by a functional $(\nn)$ such that the probability to obtain a sample with extensive distribution $\nn$ is
\begin{equation}
\label{P_biased}
   P(\nn) = \frac{W(\nn)}{r^N} 
   \left(N! \prod_{i=1}^N \frac{h_i^{n_i}}{n_i!}\right) ,
\end{equation}
where $r^N$ is a normalization constant. We require the bias to be of the form\cite{Matsoukas:E19}
\begin{equation}
\label{euler}
   \log W(\nn)
      =\sum_i n_i \frac{\partial\log W(\nn)}{\partial n_i}
      =\sum_i n_i \log w_i
\end{equation}
with $\log w_i=\partial\log W(\nn)/\partial n_i$. This makes $\log W$ homogeneous functional of $\nn$ with degree 1 and ensures that sampling converges to a limiting distribution when $N\to\infty$ \cite{Matsoukas:E19}. The log of probability to obtain the empirical distribution $p_i=n_i/N$ now is
\begin{equation}
\label{logP_biased}
   \frac{\log P(\pp )}{N} 
      = -\sum_i p_i \log \frac{p_i}{h_i w_i} - \log r .
\end{equation}
when $N\to\infty$, the set of feasible distributions contains all distributions that satisfy the normalization condition $\sum_i h_i=1$. The most probable distribution is obtained by maximizing Eq.\ (\ref{logP_biased}) with respect to $p_i$:
\begin{equation}
\label{MPD_biased}
   p^*_i = \frac{w_i}{r}h_i . 
\end{equation}
As we intended, the most probable distribution is different from the sampled distribution by a factor $w_i/r$. We can choose the partial bias $w_i$ so as to obtain \textit{any} distribution as the MPD of the biased sampling process: if we pick $w_i = r f_i/h_i$, where $f_i$ is any distribution in the domain of $h_i$, then $f_i$ will materialize as the most probable distribution by this biased sampling. In the limit $N\to\infty$ the MPD is overwhelmingly more probable than any other distribution. 

The conclusion from this analysis is this: Sampling from a distribution establishes an entire space of distributions that contains every distributions that is defined on the   domain of the sampled distribution. The bias functional establishes a probability measure on the distributions of this space. This bias can be constructed so as to select \textit{any} of these distributions as the most probable distribution. For this reason we refer to $W$ as  \textit{selection functional}. 

%--------------------------------------------------------------------------* 
\subsection{Microcanonical Sampling}
We now sample from an exponential distribution with mean $\bar x$,
\begin{equation}
   h_i = \frac{e^{-i/\bar x}}{\bar x},
\end{equation}
and consider the subspace of empirical distributions with the same mean. The probability of empirical distribution $\pp$ is given by Eq.\ (\ref{logP_biased}) with a new normalization constant $r'$. We write the result in the form
\begin{equation}
\label{logP_microcanonical}
   \frac{\log P(\pp )}{N}
   = -\sum_- p_i \log\frac{p_i}{w_i} - \log\omega .
\end{equation}
with $\log\omega= 1+\log \bar x + \log r'$. The MPD is obtained by maximizing this functional with respect to all $h_i$ that satisfy the constraint and whose mean is $\bar x$. The result is 
\begin{equation}
\label{mpd}
   p_i^* = \frac{w_i e^{-\beta i}}{ q} ,
\end{equation}
with $\beta$ and $q$ parameters associated with the Lagrange multipliers for the two constraints in this maximization.  We can show \cite{Matsoukas:E19}
\begin{equation}
\label{xav_logq_beta}
   \bar x = -\frac{d\log q}{d\beta} , 
\end{equation}
which establishes a relationship between the mean of the MPD and its parameters $\beta$ and $q$. 

%--------------------------------------------------------------------------* 
\subsection{Universal Thermodynamics}
Define the microcanonical functional $\rho[\pp ] = \log P(\pp )/N$ use Eq.\ (\ref{logP_microcanonical}) to write it as
\begin{equation}
\label{microcanonical_functional}
   \rho[\pp ] = -\sum_- p_i \log\frac{p_i}{w_i} - \log\omega . 
\end{equation}
The condition that defines the most probable distribution now is expressed in the form of the inequality $\rho[\ppp ] \geq \rho[\pp ]$. For large $N$ the MPD is overwhelmingly more probable than any other distribution. Then $P(\ppp )\to 1$ and $\rho[\ppp ]\to 0$. The condition that defines the MPD then becomes
\begin{equation}
   \rho[\pp ] \leq 0,
\end{equation}
with the equal sign only for $\pp =\ppp$. This inequality is the central variational condition of the theory. Apply this to Eq.\ (\ref{microcanonical_functional}) 
\begin{equation}
   \log\omega \geq -\sum_i h_i\log p_i + \log W[\pp ] ,
\end{equation}
or more compactly,
\begin{equation}
\label{second_law}
   \log\omega \geq S[\pp] + \log W[\pp]. 
\end{equation}
With $\pp =\ppp $ from Eq.\ (\ref{mpd}) the above becomes an exact equality. The result is
\begin{equation}
\label{fundamental}
   \log\omega = \beta \bar x + \log q .
\end{equation}
The simplest way to make contact with thermodynamics is to take $i$ to refer to the energy $E_i$ of microstate and $p_i$ to refer to the probability of microstate. The selection functional in this case is uniform, $W=w_i=1$, and expresses the postulate of equal a priori probabilities. With these substitutions (\ref{mpd}) reverts to the familiar canonical probability distribution and Eq.\ (\ref{second_law}) reads
\begin{equation}
   \log\omega \geq S[\pp ] .
\end{equation}
This is the inequality of the second law: The MPD has the maximum entropy among all possible probability distributions that can be assigned to the set of microstates. 
With Eqs.\ (\ref{mpd}), (\ref{xav_logq_beta}), (\ref{second_law}) and (\ref{fundamental}) we have obtained the generalized forms of the thermodynamic set in Eqs.\ (\ref{S_functional})--(\ref{thermo_2}) via an argument that contains no physics. In this formulation the inequality of the second law has a simple non controversial and in fact trivial meaning: it states that the most probable distribution is more probable than any other in the set of feasible distributions.

%--------------------------------------------------------------------------* 
\section{Evolution of Extensive Populations}
The implications for the universality of thermodynamics are profound. Any probability distribution in the domain of the exponential distribution $(x\geq 0)$ can be obtained as the most probable distribution in this domain and therefore can be described in the formalism of thermodynamics. The key to this formulation is the selection bias that selects the MPD among all distributions, but what is $W$? In the case of molecular systems it is fixed by the postulate of equal a priori probabilities, a \textit{model assumptions} that assigns a probability to the elements of the feasible space. In the general case it is set by the rules of the stochastic process that governs the evolution of the random variable in question. Let us sketch how this works in the case of an extensive population $\mathbf{n} = (n_1,n_2\cdots)$. We suppose the elements of the population undergo transitions of the form
\begin{equation}
   i \xrightarrow{p_{i\to j}} j
\end{equation}
with transition probability $p_{i\to j}$. Accordingly, the distribution of walkers is transformed by these transitions and produces a Markov chain of the form
\begin{equation}
   \cdots \mathbf{n'}\Big|_{G-1} 
   \to \mathbf{n}\Big|_{G} 
   \cdots
\end{equation}
where $G=0,1\cdots$ is the generation of $\mathbf{n}$ and counts the number of transitions it has undergone from fixed initial state. We then have
\begin{equation}
\label{constraints}
   G = \sum_i g_i n_i,\quad
   N = \sum_i n_i ,
\end{equation}
Here $g_i$ is the number of transitions that produce element $i$ and $N$ is the number of elements in $\mathbf{n}$ and is constant in all generations. 
%--------------------------------------------------------------------------* 
%: fig_mad
\begin{figure}
\begin{center}
\includegraphics[width=2.25in]{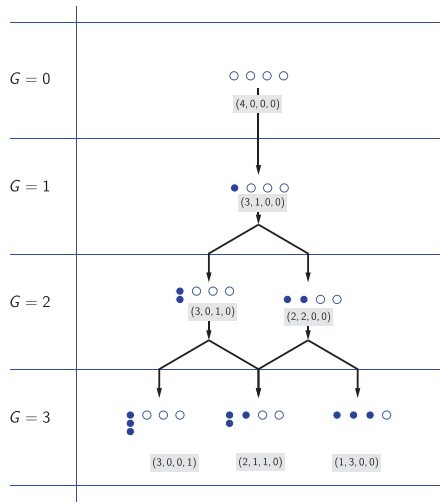}
\end{center}
\caption{Trajectories of $N=4$ walkers undergoing one transition at a time. Filled circles represent transitions accumulated by the walker. In this example $n_i$ is he number of walkers with $g_i=i$ transitions. Distributions are represented pictorially and in vector form as $(n_0,n_1,n_2,n_3)$. }
\label{fig_mad}
\end{figure}
%--------------------------------------------------------------------------* 

The ensemble $\mathcal E_{G,N}$ is the set of all distributions that are formed by $G$ sequential transitions starting from the same initial state (Fig.\ \ref{fig_mad}). The probability of distribution $\mathbf{n}$ in $\mathcal E_{G,N}$ is given by Eq.\ (\ref{P_biased}), which we now write as
\begin{equation}
\label{prob_n}
   P(\mathbf{n}) =\frac{\mathbf{n!}}{\Omega_{G,N}} W(\mathbf{n}), 
\end{equation}
with $\mathbf{n!}=N!/n_1! n_2!\cdots$ and $\Omega = \omega^N$. The distributions of the ensemble are sampled via the trajectories that lead to them and the selection functional $W$ is determined by the transition probabilities along these trajectories. The most probable distribution, obtained by maximizing Eq.\ (\ref{prob_n}) under the two constraints in (\ref{fundamental}), is given by Eq.\ (\ref{mpd}) and obeys all thermodynamic relationships in Eqs.\ (\ref{xav_logq_beta}),  (\ref{second_law}) and (\ref{fundamental}). The entire formalism of thermodynamics is transferred to generic stochastic processes. 

%--------------------------------------------------------------------------* 
\section{CONCLUSIONS}
Stripped to its core statistical thermodynamics is the calculus of the most probable distribution. The theory outlined here establishes a space of distributions (the microcanonical ensemble) and a probability measure on it via a sampling process. This mathematical construction gives rise to the relationships we recognize as thermodynamics. A stochastic process may be viewed as a network of transitions that emanate from a known initial state and branch into the future. The ensemble is the set of distributions that can be reached in a given number of steps, and sampling refers to arriving to a distribution via a trajectory of transitions. The specifics of the process (transition probabilities) enter via the selection functional --this is the contact point between generalized thermodynamics and the particulars (``physics'') of system that is being studied. We suggest that this formalism can be applied to a variety of stochastic populations and plan to offer examples  in the near future.

%--------------------------------------------------------------------------*
%: biblio
\bibliography{tm,StatMech}

\begin{thebibliography}{10}
\expandafter\ifx\csname natexlab\endcsname\relax\def\natexlab#1{#1}\fi
\expandafter\ifx\csname bibnamefont\endcsname\relax
  \def\bibnamefont#1{#1}\fi
\expandafter\ifx\csname bibfnamefont\endcsname\relax
  \def\bibfnamefont#1{#1}\fi
\expandafter\ifx\csname citenamefont\endcsname\relax
  \def\citenamefont#1{#1}\fi
\expandafter\ifx\csname url\endcsname\relax
  \def\url#1{\texttt{#1}}\fi
\expandafter\ifx\csname urlprefix\endcsname\relax\def\urlprefix{URL }\fi
\providecommand{\bibinfo}[2]{#2}
\providecommand{\eprint}[2][]{\url{#2}}

\bibitem[{\citenamefont{Gibbs}(1981)}]{Gibbs:reprint}
\bibinfo{author}{\bibfnamefont{J.~W.} \bibnamefont{Gibbs}},
  \emph{\bibinfo{title}{Elementary Principles in Statistical Mechanics}}
  (\bibinfo{publisher}{Ox Bow Press}, \bibinfo{address}{Woodbridge, CT},
  \bibinfo{year}{1981}), \bibinfo{note}{(reprint of the 1902 edition)}.

\bibitem[{\citenamefont{Shannon}(1948)}]{Shannon:BSTJ}
\bibinfo{author}{\bibfnamefont{C.~E.} \bibnamefont{Shannon}},
  \bibinfo{journal}{Bell System Technical Journal}
  \textbf{\bibinfo{volume}{27}}, \bibinfo{pages}{379} (\bibinfo{year}{1948}).

\bibitem[{\citenamefont{Jaynes}(1957)}]{Jaynes:PR57}
\bibinfo{author}{\bibfnamefont{E.~T.} \bibnamefont{Jaynes}},
  \bibinfo{journal}{Phys. Rev.} \textbf{\bibinfo{volume}{106}},
  \bibinfo{pages}{620} (\bibinfo{year}{1957}).

\bibitem[{\citenamefont{Harte et~al.}(2008)\citenamefont{Harte, Zillio,
  Conlisk, and Smith}}]{Harte:E08}
\bibinfo{author}{\bibfnamefont{J.}~\bibnamefont{Harte}},
  \bibinfo{author}{\bibfnamefont{T.}~\bibnamefont{Zillio}},
  \bibinfo{author}{\bibfnamefont{E.}~\bibnamefont{Conlisk}}, \bibnamefont{and}
  \bibinfo{author}{\bibfnamefont{A.~B.} \bibnamefont{Smith}},
  \bibinfo{journal}{Ecology} \textbf{\bibinfo{volume}{89}},
  \bibinfo{pages}{2700} (\bibinfo{year}{2008}), ISSN \bibinfo{issn}{1939-9170},
  \urlprefix\url{http://dx.doi.org/10.1890/07-1369.1}.

\bibitem[{\citenamefont{Durrett}(1999)}]{Durrett:SR99}
\bibinfo{author}{\bibfnamefont{R.}~\bibnamefont{Durrett}},
  \bibinfo{journal}{SIAM Review} \textbf{\bibinfo{volume}{41}},
  \bibinfo{pages}{677} (\bibinfo{year}{1999}),
  \eprint{https://doi.org/10.1137/S0036144599354707},
  \urlprefix\url{https://doi.org/10.1137/S0036144599354707}.

\bibitem[{\citenamefont{Timme and Lapish}(2018)}]{Timme:E18}
\bibinfo{author}{\bibfnamefont{N.~M.} \bibnamefont{Timme}} \bibnamefont{and}
  \bibinfo{author}{\bibfnamefont{C.}~\bibnamefont{Lapish}},
  \bibinfo{journal}{eNeuro}  (\bibinfo{year}{2018}),
  \eprint{http://www.eneuro.org/content/early/2018/06/29/ENEURO.0052-18.2018.full.pdf},
  \urlprefix\url{http://www.eneuro.org/content/early/2018/06/29/ENEURO.0052-18.2018}.

\bibitem[{\citenamefont{Voit}(2005)}]{Voit:05}
\bibinfo{author}{\bibfnamefont{J.}~\bibnamefont{Voit}},
  \emph{\bibinfo{title}{The Statistical Mechanics of Financial Markets}}
  (\bibinfo{publisher}{Springer Berlin Heidelberg}, \bibinfo{address}{Berlin,
  Heidelberg}, \bibinfo{year}{2005}), ISBN \bibinfo{isbn}{978-3-540-26289-3},
  \urlprefix\url{https://doi.org/10.1007/3-540-26289-X}.

\bibitem[{\citenamefont{Matsoukas}(2019{\natexlab{a}})}]{Matsoukas:springer_2019}
\bibinfo{author}{\bibfnamefont{T.}~\bibnamefont{Matsoukas}},
  \emph{\bibinfo{title}{Generalized Statistical Thermodynamics: Thermodynamics
  of Probability Distributions and Stochastic Processes}}
  (\bibinfo{publisher}{Springer International Publishing},
  \bibinfo{year}{2019}{\natexlab{a}}), ISBN \bibinfo{isbn}{978-3-030-04149-6},
  \urlprefix\url{https://doi.org/10.1007/978-3-030-04149-6}.

\bibitem[{\citenamefont{Matsoukas}(2019{\natexlab{b}})}]{Matsoukas:E19}
\bibinfo{author}{\bibfnamefont{T.}~\bibnamefont{Matsoukas}},
  \bibinfo{journal}{Entropy} \textbf{\bibinfo{volume}{21}}
  (\bibinfo{year}{2019}{\natexlab{b}}), ISSN \bibinfo{issn}{1099-4300},
  \urlprefix\url{https://www.mdpi.com/1099-4300/21/9/890}.

\bibitem[{\citenamefont{Kullback and Leibler}(1951)}]{Kullback:AMS51}
\bibinfo{author}{\bibfnamefont{S.}~\bibnamefont{Kullback}} \bibnamefont{and}
  \bibinfo{author}{\bibfnamefont{R.~A.} \bibnamefont{Leibler}},
  \bibinfo{journal}{Ann. Math. Statist.} \textbf{\bibinfo{volume}{22}},
  \bibinfo{pages}{79} (\bibinfo{year}{1951}),
  \urlprefix\url{http://dx.doi.org/10.1214/aoms/1177729694}.

\end{thebibliography}
\bibliographystyle{apsrev}
%--------------------------------------------------------------------------*
\end{document}